\documentclass[aps, 12pt, final, notitlepage, oneside, onecolumn, nobibnotes,
nofootinbib, noshowpacs,amsmath,amssymb] 
{revtex4}
\usepackage{bm}
\usepackage{epsfig}
\usepackage{graphics}

\usepackage{graphicx}
\usepackage{dcolumn}
\usepackage{bm}
\usepackage[english]{babel}

\begin{document}

\title{Viscosity of Cobalt Melt: Experiment, Simulation, and Theory}

\author{R.M. Khusnutdinoff$^{a,*}$, A.V. Mokshin$^{a,**}$, A.L. Bel'tyukov$^{b}$, and N.V. Olyanina$^{b}$}
\affiliation{$^{a}$Kazan Federal University, Kazan, Russia\\
$^{b}$Physical-Technical Institute, Ural Branch of the Russia Academy of Sciences, Izhevsk, Russia\\
*e-mail: khrm@mail.ru\\
**e-mail: anatolii.mokshin@mail.ru}

\begin{abstract}
\textbf{Abstract} $-$ The results of experimental measurements, molecular dynamics simulation, and theoretical calculations of the viscosity of a cobalt melt in a temperature range of $1400-2000$~K at a pressure $p=1.5$~bar corresponding to an overcooled melt at temperatures of $1400-1768$~K and an equilibrium melt with temperatures from the range $1768-2000$~K are presented. Theoretical expressions for the spectral density of the time-dependent correlation function of the stress tensor $\tilde{S}(\omega)$ and kinematic viscosity $\nu$ determined from the frequency and thermodynamic parameters of the system are obtained. The temperature dependences of the kinematic viscosity for the cobalt melt are determined experimentally by the torsional oscillation method; numerically, based on molecular simulation data with the EAM potential via subsequent analysis of the time correlation functions of the transverse current in the framework of generalized hydrodynamics; and by the integral Kubo-Green relation; they were also determined theoretically with the Zwanzig-Mori memory functions formalism using a self-consistent approach. Good agreement was found between the results of theoretical calculations for the temperature dependence of the kinematic viscosity of the cobalt melt using experimental data and the molecular dynamics simulation results. From an analysis of the temperature dependence of the viscosity, we obtain an activation energy of $E=(5.38\pm0.02)\times10^{-20}$~J.
\end{abstract}
\pacs{\textbf{DOI}: 10.1134/S0018151X18020128 }

\maketitle

\section{Introduction}

Amorphous-forming fluids that do not crystallize upon cooling and retain a disordered structure down to very low temperatures are a subject of intensive research in condensed matter physics \cite{1,2,3,4}. A specific feature of such systems is associated with the temperature dependence of the viscosity (or the structural relaxation time), which varies by more than 15 orders of magnitude when passing from the liquid to the amorphous phase \cite{5}. Amorphous metallic alloys (AMAs) are of particular interest, since they have unique physicomechanical properties \cite{6,7,8,9,10}. Typically, AMAs are a multicomponent system with a high glass-forming ability that forms an amorphous phase at cooling with the rates $\gamma = (10^4 -10^7)$~K/s \cite{11,12}. As shown in numerous molecular dynamics studies \cite{13,14,15,16,17,18,19}, an amorphous phase can also be obtained in the case of one-component (pure) metals as a result of superfast quenching $(\gamma = 10^{11}-10^{13}$~K/s). It is noteworthy that ferromagnetic transition metals (Fe, Ni, and Co), which are widely used in the aerospace industry, represent a particular case of single-component glassforming metallic systems \cite{7,8,20}. In this case, cobalt, as compared with iron and nickel, remains poorly understood. Thus, in particular, problems related to transport processes (self-diffusion, viscosity, thermal conductivity, and electrical conductivity) and the mechanisms by which collective excitations propagate in an equilibrium liquid and supercooled cobalt phases are unclear. This is partially due to the lack of experimental viscosimetry, inelastic X-ray, and neutron scattering data for this substance \cite{20,21}. The known experimental data on the viscosity of equilibrium liquid cobalt obtained by different researchers \cite{22} reveal a significant (above $30\%$) discrepancy, which indicates the need for further research to clarify the absolute values of viscosity. Note that no viscosity values of the supercooled cobalt melt are currently found in the scientific literature.

This paper discusses the results of experimental measurements and molecular dynamics calculations of the viscosity of a cobalt melt in equilibrium liquid and supercooled phases. The kinematic viscosity was determined for the domain above and below the melting point $T_m(Co)=1768$~K: in the experiment for the temperature range $T = 1506-1969$~K and in simulation for $1400-2000$~K.

\section{Experimental} \label{Experimental}

Experimental measurements were carried out for cobalt metal grade K0, which has a weight fraction of cobalt of at least $99.98\%$ and contains the following impurities: $0.003\%$ of Fe; less than $0.005\%$ of Ni and C;
$0.001\%$ of Si, Cu, Mg, Zn, and Al; and $<0.001\%$ of O. The kinematic viscosity $\nu$ was measured by the method of torsional vibrations of a cylindrical crucible with a melt in the Shvidkovskii variant \cite{23} with an automated set-up \cite{24} in a protective helium atmosphere. Cylindrical $Al_2O_3$ cups with an internal diameter of $\sim17$~mm and a height of $\sim42$~mm were used as crucibles. To prevent the uncontrolled influence of the oxide film that is formed on the surface of the alloy during the measurement process, an $Al_2O_3$ lid was placed over the sample in the crucible. The lid construction is described in \cite{25}. The gap between the side walls of the crucible and the lid was $0.2-0.3$~mm. During the measurements, the lid tightly adhered to the upper boundary of the melt, providing a reliable friction surface. Rotation of the lid relative to the crucible was not possible.

The temperature dependences of the viscosity were measured in a heating regime from the melting point of cobalt to $1973$~K and subsequent cooling to its crystallization in steps of $15-25$~K. At each temperature, the melt was held for $20$~min, after which no less than ten measurements were performed. The temperature of the melt was determined with an accuracy of $\pm 5$~K with a tungsten-rhenium thermocouple, $3-4$~mm under the bottom of the crucible, calibrated for the melting points of pure metals (Al, Cu, Ni, Fe).

The kinematic viscosity values were calculated by numerical solving the equation of motion of the cup \cite{23},\cite{24}:

\begin{displaymath}
\textrm{Re}(L)+\dfrac{\delta}{2\pi}\textrm{Im}(L)-2I \left(\dfrac{\delta}{\tau}- \dfrac{\delta_0}{\tau_0} \right) = 0,
\end{displaymath}
where $I$ is the moment of inertia of the suspension system; $\delta,~ \tau,~ \delta_0,~ \tau_0$ are the damping decrement and the period of oscillations of the suspension system with and without a melt, respectively; $\textrm{Re}(L)$ and $\textrm{Im}(L)$ and the real and imaginary parts of the friction function, taking into account two side surfaces of friction.

The thermal expansion of the crucible material entered the value of the sample radius $R$, taking into account that, in the range from $273$~K to $2073$~K, the mean linear thermal expansion coefficient of $Al_2O_3$ is $9.0 \times 10^{-6}~\textrm{deg}^{-1}$ \cite{26}. The height of the melt in the crucible was calculated by the equation
\begin{displaymath}
H = \dfrac{m}{\pi R^2\rho},
\end{displaymath}
where $m$ is the sample weight and $\rho$ is the density of the cobalt melt, $kg/m^3$. The density of liquid cobalt was calculated by the equation obtained \cite{22} by averaging the data from various authors:

\begin{displaymath}
\rho = 6172.152 - 0.936 T.
\end{displaymath}

To estimate the errors in measuring viscosity, the method described in \cite{27}, was used. It was established that the relative error of the obtained kinematic viscosity values does not exceed $4\%$, with an error in a single experiment of $2\%$.

\section{Simulation} \label{Simulation}

The molecular dynamics simulations of the cobalt melt were performed in an isothermally isobaric $(NpT)$ ensemble in a temperature range of $1400-2000$~K and at $p = 1.5$~bar. The system consisted of $N = 4000$ atoms in a cubic cell with periodic boundary conditions. The interaction between atoms was taken into account by the EAM potential \cite{28}
\begin{align*}
U=\sum \varphi (r_{ij})+\sum F(\rho_i),\\
\rho_i=\sum \psi (r_{ij}).\qquad
\end{align*}

Here, $\varphi(r)$ is the pair potential of interparticle interaction, and $F(\rho)$ is the embedded function, which effectively takes into account multiparticle correlations via the electron density $\rho_i$ of the $i$th atom. The supercooled cobalt melt was obtained by rapid cooling of the equilibrium melt (at $2000$~K) at a cooling rate of $\gamma=10^{12}$~K/s \cite{29}. Integration of the equations of motion was carried out according to the Verlet velocity algorithm with a time step of $\tau = 10^{-15}$~s \cite{30}. To bring the system to a state of thermodynamic equilibrium and to calculate the spectral characteristics at each temperature of $1400,~ 1500,~ 1600,~ 1700,~ 1800,~ 1900$ and $2000$~K, the program performed $10^5$ and $5 \times 10^6$ time steps, respectively.

The simplest way to verify the accuracy of the potential of interparticle interaction to reproduce of the structural properties of the system is to calculate the radial distribution function of atoms \cite{31}

\begin{displaymath}
g(r)=\dfrac{V}{N^2} \left\langle \sum_i \sum_{j\neq i}\delta (r-r_{ij}) \right\rangle ,
\end{displaymath}
and the statistic structure factor \cite{32},\cite{33}
\begin{displaymath}
S(k)=1+\dfrac{N}{V} \int [g(r)-1]\exp[i(\textbf{k,r})]d \bf{r},
\end{displaymath}
which can be compared with the experimental X-ray and neutron diffraction data.

\begin{figure}[tbh]
\includegraphics[height=8.5cm, angle=0]{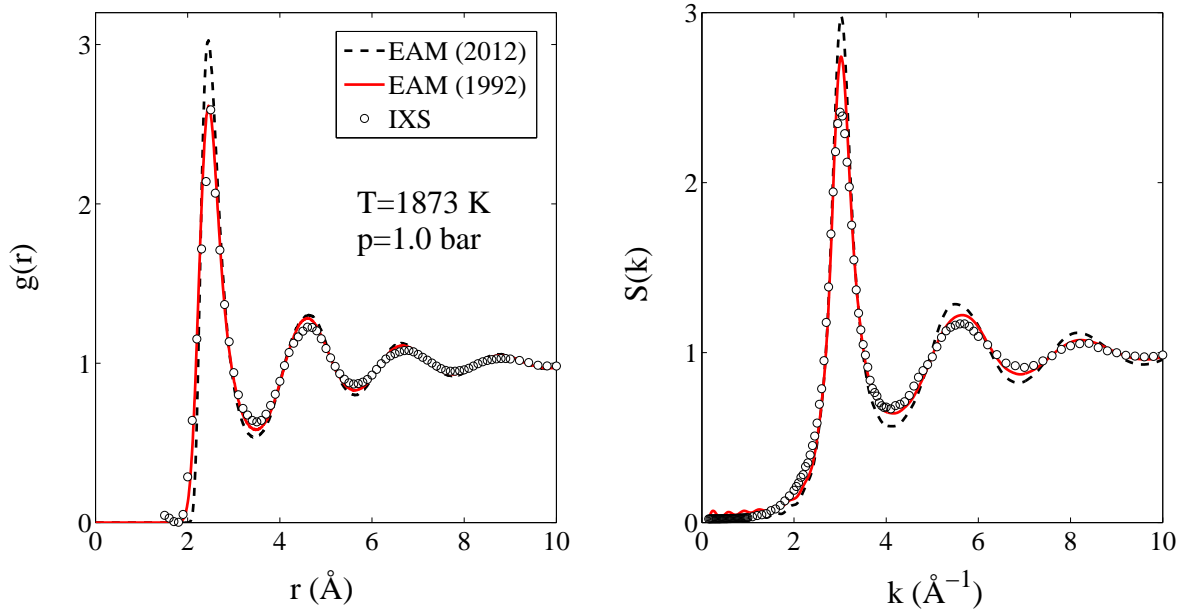}
\caption{Radial distribution function of cobalt atoms at $T=1873$~K (a) and the static structure factor of the cobalt melt (b): curves -- the results of molecular dynamics simulation obtained on the basis of the interatomic interaction model \cite{34} (\ref{1}) and the potential \cite{28} (\ref{2}); (\ref{3}) experimental data on X-ray diffraction \cite{7}.}
\label{one}
\end{figure}

Figure 1 shows the radial distribution function $g(r)$ of atoms and the static structure factor $S(k)$ of liquid cobalt at $T = 1873$~K. The simulation results are compared with the experimental X-ray diffraction data \cite{7}. Calculation with the EAM potential \cite{28} yields better agreement with the experimental data compared with the results on the basis of the model potential from \cite{34}: the $S(k)$ components and the full form of the radial distribution function are correctly reproduced\footnote{Unfortunately, the authors do not know the experimental data for $S(k)$ of liquid cobalt at other temperatures.}. This result is somewhat unexpected, since the potential, proposed by Pan and Mishin \cite{34}, refers to the ``new-generation'' EAM potentials; therefore, we could quite naturally assume that it reproduces the structural characteristics of the melt more qualitatively.

Calculation of $\nu$ using simulation data can be done in several ways. Firstly, it is possible to determine this quantity by the generalized Einstein relation \cite{35,36}
\begin{displaymath}
\nu =\dfrac{m}{k_BTN}\lim_{t\rightarrow\infty}\dfrac{1}{2t} \left\langle \left\vert \sum_{i=1}^N [\upsilon_{ix}(t)r_{iy}(t)-\upsilon_{ix}(0)r_{iy}(0)]\right \vert^2 \right\rangle.
\end{displaymath}

Second, the coefficient $\nu$ can be found by the Kubo-Green integral relation \cite{37}
\begin{displaymath}
\nu =\dfrac{VS_0}{\rho k_BT} \int_0^\infty S(t)dt,
\end{displaymath}
where $S(t)$ is the normalized time correlation function of the stress tensor (the expression for $S(t)$ is presented below) and $k_B$ is the Boltzmann constant.

Lastly, to determine $\nu$, it is possible to use the conclusions of generalized hydrodynamics \cite{37}, according to which it is assumed that the generalized kinematic viscosity coefficient $\nu(k, \omega)$ is considered as
\begin{displaymath}
\nu =\lim_{\omega\rightarrow 0} \lim_{k\rightarrow 0} \nu(k,\omega).
\end{displaymath}

Thus, from the well-known hydrodynamics equation
\begin{displaymath}
\dfrac{\partial}{\partial t}C_T(k,t)+\nu k^2 C_T(k,t)=0
\end{displaymath}
we obtain \cite{38,39}
\begin{eqnarray}
\lim_{\omega\rightarrow 0}\nu(k,\omega)= \left( k^2\int_0^\infty C_T(k,t)dt \right)^{-1},
\label{1}
\end{eqnarray}
where $C_T(k, t)$ is the normalized time correlation function of the transverse current \cite{40}:
\begin{displaymath}
C_T(k, t)= \dfrac{\langle [\textbf{e}_k,\textbf{j}^{\ast}(k,0)],[\textbf{e}_k,\textbf{j}(k,t)] \rangle}{\langle \vert[\textbf{e}_k,\textbf{j}(k,0)]\vert^2\rangle}.
\end{displaymath}

Here, the scalar and vector products are denoted by parentheses and brackets, respectively; $\textbf{j}(k,t)$ is the microscopic current expressed by the equation \cite{37,41}
\begin{displaymath}
\textbf{j}(k,t)= \dfrac{1}{\sqrt{N}}\sum_l^N \textbf{v(t)}\exp[-i(\textbf{k,r}_l(t))],
\end{displaymath}
where $\textbf{v}_l(t)$ is the velocity of $l$ th particle at time instant $t$ and $\textbf{e}_k =\textbf{k}/\vert\textbf{k}\vert$ is the unit vector codirectional with wave vector \textbf{k}. Conversely, in a long-wavelength limit, the following expression takes place \cite{42}:

\begin{eqnarray}
\lim_{k\rightarrow 0}\nu(k,0)= \dfrac{\nu}{1+\alpha^2k^2}.
\label{2}
\end{eqnarray}

Interpolating the dependence on $k$ for the fixed temperature by means of (\ref{2}), we can determine the value of the coefficient $\nu$.

\section{Theoretical formalism} \label{Theoretical formalism}

Let us consider a system consisting of $N$ identical particles of mass $m$ in volume $V$. We choose the off-diagonal components of the stress tensor \cite{43} as the initial dynamic variable
\begin{displaymath}
\sigma_{\alpha,\beta}=\dfrac{1}{V}\left(\sum_{i=1}^N mv_{i\alpha}v_{i\beta}-\sum_{i=1}^{N-1}\sum_{j=i+1}^N r_{ij\alpha}\dfrac{\partial U(r_{ij})}{\partial r_{ij\beta}}\right),
\end{displaymath}
where $r_{ij}=\vert\textbf{r}_i-\textbf{r}_j\vert$ is the distance between the particles with indices $i$ and $j$; $\textbf{v}_i$ is the vector of particle velocity $i$; and $U(r_{ij})$ is the potential of interparticle interaction. Indices $\alpha ,\beta \in \{x, y, z\}$ denote the projection of values on the corresponding coordinate axis. We determine the time correlation function (TCF) of stress tensor \cite{44,45}
\begin{displaymath}
S(t)=\dfrac{\langle\sigma_{\alpha,\beta}(t)\sigma_{\alpha,\beta}(0)\rangle}{\langle\vert\sigma_{\alpha,\beta}(0)\vert^2\rangle}
\end{displaymath}
where
\begin{align*}
S_0=\langle\vert\sigma_{\alpha,\beta}(0)\vert^2\rangle =\left( \dfrac{k_BT}{V}\right)+\dfrac{2\pi n}{15}\dfrac{k_BT}{V^2}\int_0^\infty r^4 g(r)(5B+Ar^2)dr,
\end{align*}
\begin{displaymath}
B=\dfrac{1}{r}\dfrac{\partial U(r)}{\partial r}, \quad A=\dfrac{1}{r}\dfrac{\partial B(r)}{\partial r}
\end{displaymath}

The short-time expansion of $S(t)$ can be presented in the form \cite{46}
\begin{displaymath}
S(t)=1-S^{(2)}\dfrac{t^2}{2!}+S^{(4)}\dfrac{t^4}{4!}-S^{(6)}\dfrac{t^6}{6!}+ \cdots
\end{displaymath}

Here, $S^{(2m)}$ is the even frequency moments \footnote{The microscopic expressions for the frequency moments of $S^{(2)}$ and $S^{(4)}$ are presented in \cite{47}.}
\begin{equation}
S^{(2m)}=\dfrac{\int \omega^{2m} \tilde{S}(\omega)d\omega}{\int \tilde{S}(\omega)d\omega}, \quad m=1,2,\cdots
\label{3}
\end{equation}
of the spectral density of TCF of the stress tensor \cite{20}
\begin{displaymath}
\tilde{S}(\omega)=\dfrac{S(t=0)}{2\pi}\textrm{Re} \int_{-\infty}^\infty \exp(i\omega t)S(t)dt.
\end{displaymath}

Conversely, the spectral density of the TCF of the stress tensor $\tilde{S}(\omega)$ can be represented in the form \cite{48}
\begin{displaymath}
\tilde{S}(\omega)=\dfrac{S(t=0)}{\pi}\textrm{Re} \left\lbrace\dfrac{1}{-i\omega +\Delta_1 \tilde{M}(\omega)} \right\rbrace ,
\end{displaymath}
where $\tilde{M}(\omega)$ is the spectral density of so-called first-order memory function, which is connected with the high-order memory functions $\tilde{M}_n(\omega)$ at $n > 1$ by the recurrence relation\footnote{We note that in the time domain this relation can be represented as an infinite chain of integro-differential Zwanzig-Mori kinetic equations \cite{49,50}.}
\begin{displaymath}
\tilde{M}_n(\omega)= [-i\omega +\Delta_{n+1}\tilde{M}_{n+1}(\omega)]^{-1},
\end{displaymath}
and $\Delta_n$ are the frequency parameters, which are expressed by the spectral moments $S^{(2m)}, n, m = 1, 2, \ldots$:
\begin{align*}
\Delta_1=S^{(2)},\qquad\\
\Delta_2=\dfrac{S^{(4)}}{S^{(2)}}-S^{(2)},\quad\\
\Delta_2=\dfrac{S^{(6)}S^{(2)}-S^{(4)^2}}{S^{(4)}S^{(2)}-S^{(2)^3}},\\
\cdots \qquad \qquad
\end{align*}

In accordance with the Kubo-Green formula for kinematic viscosity $\nu$, we have \cite{20}
\begin{equation}
\nu=\dfrac{VS_0}{\rho k_BT}\int_0^{\infty}S(t)dt=\dfrac{VS_0}{\rho k_BT} \tilde{S}(\omega=0).
\label{4}
\end{equation}

As shown in \cite{51,52}, the condition $\tilde{M}_2(\omega) = \tilde{M}_1(\omega)$ is realized for transport processes in single-component liquids, which allows us to find an expression for the spectral density
\begin{equation}
\tilde{S}(\omega)=\dfrac{1}{\pi}\dfrac{2\Delta_1\Delta_2\sqrt{4\Delta_2-\omega^2}}{\Delta_1^2(4\Delta_2-\omega^2)+\omega^2(2\Delta_2-\Delta_1)^2}.
\label{5}
\end{equation}

From this, we obtain a simple expression for the kinematic viscosity
\begin{equation}
\nu=\dfrac{VS_0}{\pi\rho k_BT}\dfrac{\sqrt{\Delta}_2}{\Delta_1}.
\label{6}
\end{equation}

Here, the frequency parameter $\Delta_1$ is determined by an integral expression containing $U(\textbf{r})$ and the radial distribution functions of two and three particles, $g(\textbf{r})$ and $g_3(\textbf{r}_1, \textbf{r}_2)$, respectively. For numerical estimation of parameter $\Delta_2$, it is also necessary to know the distribution function of four particles $g_4(\textbf{r}_1, \textbf{r}_2, \textbf{r}_3)$ (see expressions (\ref{5}) and (\ref{6}) in \cite{47}).

\section{Results} \label{Results}

Figure 2 shows a comparison of the results of theoretical calculations for the spectral density $\tilde{S}(\omega)$ of the cobalt melt by Eq.\ref{5} with molecular dynamics simulation data. The frequency moments of the spectral
density of the stress tensor TCF $\tilde{S}(\omega)$ used in the calculations are presented in Table 1. The values of the moment $S^{(0)}$ were determined from the condition for normalizing the TCF of the stress tensor $S(t = 0) = 1$, the frequency moment $S^{(2)}$ was calculated by numerical integration of the simulation spectra in accordance with the relation (\ref{3}), and the fourth moment $S^{(4)}$ was found by comparing the theory with the simulation results. The theoretical curves correctly reproduce all of the features of the $\tilde{S}(\omega)$ spectra at all of the considered temperatures, which indicates indirectly the correctness of the theoretical model (\ref{6}). 
\begin{figure}[tbh]
\includegraphics[height=8.5cm, angle=0]{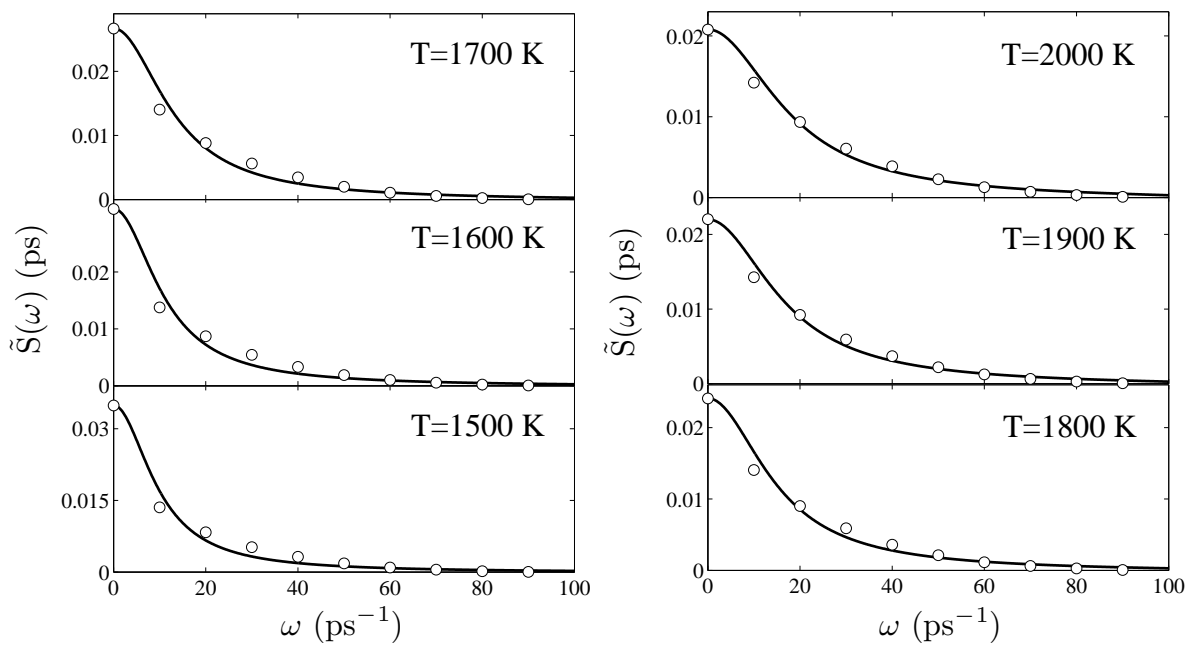}
\caption{Spectral density of the time correlation function of the stress tensor $\tilde{S}(\omega)$ of the cobalt melt at various temperatures: curves stand for the results of theoretical calculations according to Eq. (\ref{5}), dots -- molecular dynamics simulation data.}
\label{two}
\end{figure}
At the same time, comparison with experimental data (for example, inelastic scattering of neutrons or X-rays) is required in order to unambiguously conclude that the EAM potential from \cite{28} is able to reproduce correctly the microscopic dynamics of the cobalt melt in the considered temperature range. At the moment, such experimental data are not available.

\begin{center}
\textbf{Table 1.} Frequency moments $S^{(2m)}$ of the spectral density TCF of the stress tensor (\ref{5}) and the fitting parameters to the expression (\ref{2}) for the cobalt melt at $p = 1.5$~bar
\begin{tabular}{|c|c|c|c|c|} \hline 
~~~$T$, K~~~ &$~~S_0$, $10^{16}$ \textrm{Pa}$^2$~~ & $~~ S^{(2)}$, $10^{26}$ s$^{-2}$~~ & ~~$S^{(4)}$, $10^{54}$ s$^{-4}$~~ & ~~ $\alpha$, $\textrm{\AA}$~~ \\\hline
1400 & 2.107 & 6.455 & 4.858 & 0.54\\\hline
1500 & 2.162 & 6.804 & 4.251 & 0.54\\\hline
1600 & 2.203 & 7.120 & 3.953 & 0.54\\\hline
1700 & 2.228 & 7.479 & 3.504 & 0.54\\\hline
1800 & 2.248 & 7.876 & 3.422 & 0.55\\\hline
1900 & 2.260 & 8.162 & 3.263 & 0.58\\\hline
2000 & 2.274 & 8.484 & 3.228 & 0.60\\\hline
\end{tabular}
\end{center}
\begin{figure}[tbh]
\includegraphics[height=8.5cm, angle=0]{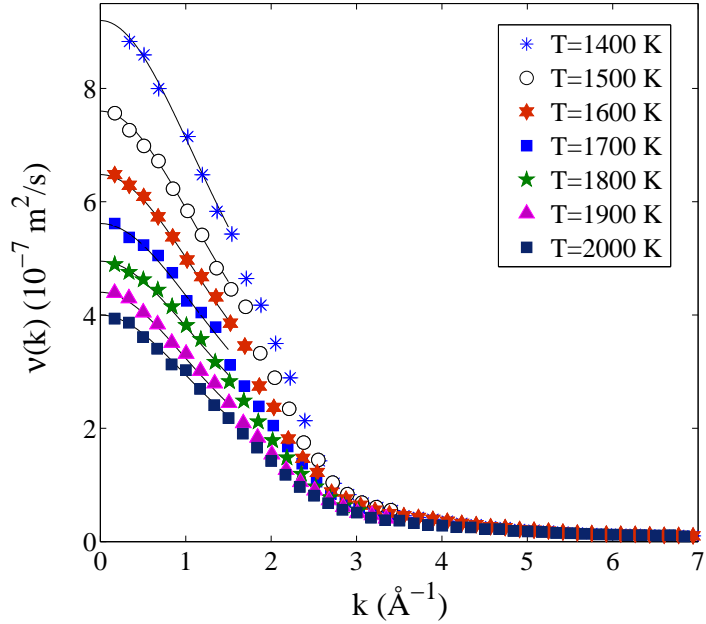}
\caption{Wave number dependencies of the kinematic viscosity of the cobalt melt at $p=1.5$~bar and various temperatures, obtained on the basis of simulation data: (\ref{1}) $T=1400$~K, (\ref{2}) $1500$, (\ref{3}) $1600$, (\ref{4}) $1700$, (\ref{5}) $1800$, (\ref{6}) $1900$, (\ref{7}) $2000$; solid lines show the results of the fitting procedure with (\ref{2}).}
\label{three}
\end{figure}

Figure 3 shows the simulation results for the wave number dependence of the kinematic viscosity at various temperatures. The values $\nu$ were directly determined by the interpolation of these curves into the region of small values of $k$. Table 1 presents the parameters obtained as a result of fitting expression (\ref{2}) to $\nu(k, 0)$ from the simulation. Figure 4 shows the temperature dependence of the kinematic viscosity for the cobalt melt at the pressure $p = 1.5$ bar. The results of theoretical calculations (\ref{6}) for the temperature dependence of the kinematic viscosity of the cobalt melt are in good agreement with both the molecular dynamics simulation results and with the experimental data. We note that the obtained results and experimental results have minor discrepancies as compared to the experimental data of previous measurements \cite{22,53}.

\begin{figure}[tbh]
\includegraphics[height=8.5cm, angle=0]{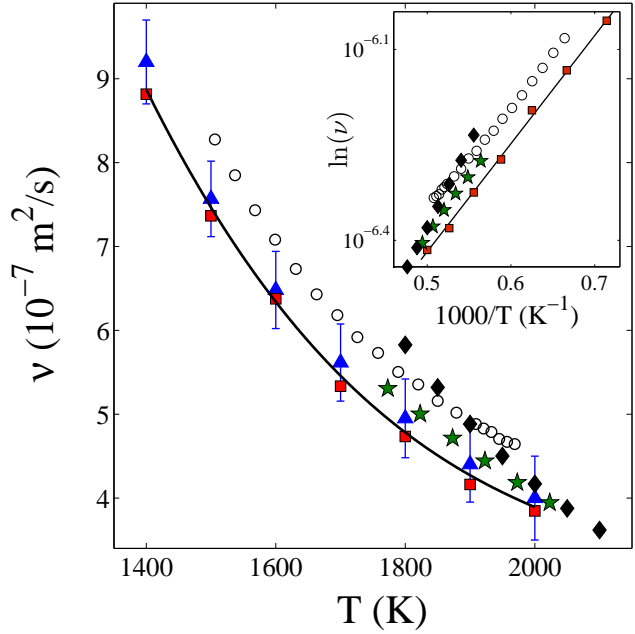}
\caption{The temperature dependence of the kinematic viscosity of the cobalt melt at $p=1.5$~bar: (\ref{1}),(\ref{2}) results of molecular dynamics simulation calculated with the Kubo-Green relation for $S(t)$ (formula \ref{4}) and in the framework of generalized hydrodynamics with $C_T(k, t)$ (Eqs. (\ref{1}) and (\ref{2})), respectively; (\ref{3}) experimental results; solid line -- the results of theoretical calculations performed according to (\ref{6}); (\ref{4}) and (\ref{5}) experimental data from \cite{22} and \cite{53}, respectively; on the inset -- the dependence in a logarithmic scale: (\ref{6}), (\ref{7}) the results of simulation and experimental results, respectively; solid line -- results of fitting with (\ref{7}).}
\label{four}
\end{figure}

As is known, for equilibrium liquids, the temperature dependence of the viscosity should be determined by the Arrhenius thermal activation law
\begin{equation}
\nu (T)=\nu_0 \exp\left(\dfrac{E}{k_BT}\right),
\label{7}
\end{equation}
where $\nu_0$ is the preexponential factor corresponding to the formal value of viscosity at $T\rightarrow\infty$ and $E$ is the height (energy) of activation barrier of viscosity process. The dependence of viscosity $\nu(T)$ on the temperature is presented in Fig. 4 in a logarithmic scale. Fitting by Eq. (\ref{7}) was performed with the parameters
$\nu_0^{MD} = (5.4 \pm 0.05) \times 10^{-8}$~m$^2/$s ($\nu_0^{Exp}=(6.3\pm 0.1)\times 10^{-8}$~m$^2/$s) and $E^{MD} = (5.4 \pm 0.02) \times 10^{-20}$~J ($E^{Exp} = (5.35 \pm 0.05) \times 10^{-20}$~J) for simulation and experiment, respectively. It can be seen that the results of the present experiment and the simulation for the temperature dependence of the kinematic viscosity are reproduced well by the Arrhenius relation (\ref{7}). It should be noted that the experimental values at $T<1631$~K begin to deviate insignificantly from this dependence. At the same time, the experimental data \cite{53} are closer to the results of simulation and have a smaller scattering in comparison with the experimental results of this work. The experimental values of the viscosity from \cite{22} cover the high-temperature region of $1800-2100$~K. Although these data approximate the results of numerical calculations at high temperatures, the character of the temperature dependence of $\nu(T)$ in accordance with these experimental data differs both from other experiments and simulation results. Thus, we can estimate the values of $\stackrel{-}{\nu}_0$ and $E$, and thus obtain $\stackrel{-}{\nu}_0=(5.8 \times 0.1)\times 10^{-8}$~m$^2/$s and $\stackrel{-}{E}=(5.38\pm 0.02) \times 10^{-20}$~J.

\section{Conclusions} \label{Conclusions}

The viscosity properties of cobalt melt in a temperature range of $1400-2000$~K at $p=1.5$~bar were investigated experimentally by the torsional vibration method and numerically by molecular dynamics simulation with the EAM potential of interparticle interaction \cite{28}. To verify the accuracy of the considered EAM potential to reproduce the particle dynamics in the cobalt melt, the structural characteristics of the system (the radial distribution function and the static structure factor) were calculated. They show good agreement with the experimental data on X-ray diffraction \cite{7}. Calculation of the kinematic viscosity on the basis of the results of molecular dynamics simulation was carried out in two ways: with the Kubo-Green integral relation for $S(t)$ and within the framework of generalized hydrodynamics with TCF of the transverse current $C_T(k,t)$. It is shown that both the methods lead to similar temperature dependencies of the kinematic viscosity, which are in good agreement with the experimental results of viscosimetry.

Within the formalism of memory functions and the self-consistent approach \cite{51,52}, expressions were obtained for the spectral density of the stress tensor time correlation function  $\tilde{S}(\omega)$ and kinematic viscosity, which are determined from the frequency parameters $\Delta_1$ and $\Delta_2$. The results of theoretical calculations of the temperature dependence of the viscosity $\nu(T)$ of the cobalt melt are in good agreement with both the results of molecular dynamics simulations and with the experimental data.

\section{Acknowledgments} \label{Acknowledgments}

Large-scale molecular dynamics calculations were performed on the computing cluster of the Kazan Federal University and the supercomputer of the Interdepartmental Supercomputer Center of the Russian Academy of Sciences. This work was supported by the Ministry of Eduction and Science of the Russian Federation via Kazan State University in the framework of the task 3.2166.2017/4.6 and by the grant from the President of the Russian Federation MD-5792.2016.2.

\end{document}